\documentclass[iop]{emulateapj}

\usepackage{graphicx}

\shorttitle{Imbalanced Relativistic MHD Turbulence}
\shortauthors{Cho \&  Lazarian}

\begin{document}

\title{Imbalanced Relativistic Force-Free Magnetohydrodynamic Turbulence} 
\author{Jungyeon Cho\altaffilmark{1} and A. Lazarian\altaffilmark{2}}
\altaffiltext{1}{Department of Astronomy and Space Science,
       Chungnam National University, Daejeon, Korea; jcho@cnu.ac.kr} 
\altaffiltext{2}{Department of Astronomy, University of Wisconsin,
       Madison, WI 53706, USA}


\begin{abstract}
When magnetic energy density is much larger than that of matter, 
as in pulsar/black hole magnetospheres,   
the medium becomes force-free and we need relativity to describe it. 
As in non-relativistic magnetohydrodynamics (MHD), 
Alfv\'enic MHD turbulence in the relativistic limit 
can be described by interactions of counter-traveling
wave packets.
In this paper we numerically study strong imbalanced MHD
turbulence in such environments.
Here, imbalanced turbulence
means the waves traveling
in one direction (dominant waves) have higher amplitudes 
than the opposite-traveling waves (sub-dominant waves).
We find that (1) spectrum of the dominant waves is steeper than that of sub-dominant waves, (2)
 the anisotropy of  the dominant waves is weaker than that of sub-dominant waves, and
(3) the dependence of the ratio of magnetic energy densities of dominant and sub-dominant waves 
on the ratio of energy injection rates is steeper than quadratic
 (i.e.,~$b_+^2/b_-^2 \propto (\epsilon_+/\epsilon_-)^n$ with $n>2$).
These results are consistent with those obtained for imbalanced non-relativistic Alfv\'enic turbulence.
This corresponds well to the earlier reported similarity of the relativistic and non-relativistic
balanced magnetic turbulence.
\end{abstract}
\maketitle

\section{Introduction}
Alfv\'en waves play important roles in strongly magnetized media.
They propagate along magnetic field lines with the Alfv\'en speed $V_A \equiv B_0/\sqrt{4 \pi \rho}$,
where $B_0$ is the strength of the mean magnetic field and $\rho$ is density.
Alfv\'en waves moving in opposite directions can interact and result in
Alfv\'enic magnetohydrodynamic (MHD) turbulence.

Alfv\'enic MHD turbulence in the non-relativistic limit has been studied for many decades and the
best available MHD turbulence model is, 
in spite of all existing controversies 
\citep[see][]{MarG01,Muletal03,Bol05,BerL06,Matt08,Cho10,Ber11},
 the one by 
Goldreich \& Srdihar (1995; henceforth GS95)
which was first numerically tested by Cho \& Vishniac (2000). 
The GS95 model predicts a Kolmogorov spectrum 
($E(k)\sim k^{-5/3}$) and scale-dependent anisotropy ($k_{\|} \propto k_{\perp}^{2/3}$),
where $k_{\|}$ and $k_{\perp}$ are wave-numbers along and perpendicular to
the local mean magnetic field directions, respectively, and 
$k=\sqrt{ k_\perp^2+k_\|^2 }$.

When $B_0$ goes to infinity and/or $\rho$ goes to zero, 
Alfv\'en speed approaches the speed of light and a new regime
of turbulence emerges.
More precisely, when the magnetic energy density is so large that
the inertia of the charge carriers can be neglected,
the medium can be described by relativistic force-free MHD equations \citep{GolJ69,BlaZ77,ThoB98}.
Cho (2005) numerically studied  three-dimensional
MHD turbulence in this
extreme relativistic limit 
and found the following results.
First, the energy spectrum is consistent with a Kolmogorov spectrum:
$E(k)\sim k^{-5/3}$.
Second, turbulence shows the Goldreich-Sridhar type
anisotropy: $k_{\|} \propto k_{\perp}^{2/3}$.
These scaling relations are in agreement with earlier theoretical predictions
by Thompson \& Blaes (1998).

The similarity between non-relativistic Alfv\'enic MHD turbulence 
and relativistic force-free MHD turbulence 
leads us to the question: 
to what extent are
relativistic and non-relativistic 
Alfv\'enic turbulence similar?
In this paper, we try to answer this question.
Strong imbalanced Alfv\'enic turbulence is an ideal problem for that purpose 
because interactions between eddies
are very complicated in strong imbalanced Alfv\'enic turbulence.  
In imbalanced Alfv\'enic turbulence, 
the waves traveling
in one direction (dominant waves) have higher amplitudes 
than the opposite-traveling waves (sub-dominant waves).
By `strong' imbalanced turbulence, we mean the dominant waves satisfy
the condition of critical balance, $b k_{\perp} /(B_0 k_{\|} ) \sim 1$, 
at the energy injection scale, where $b$ is the strength of the fluctuating magnetic field.

Many studies exist for strong imbalanced Alfv\'enic turbulence in the non-relativistic limit 
\citep{LitGS07,BerL08,Chan08,BerL09,PerB09,PodB10,Per12,MasBC12}, 
but no study is available yet for its relativistic counterpart.
In this paper, we
compare our relativistic simulations with non-relativistic ones.
Our study can have many astrophysical implications.
So far, we do not fully understand turbulence processes in extremely relativistic environments, such as
black hole/pulsar magnetospheres, or gamma-ray bursts.
If we can verify close similarities between extremely relativistic and Newtonian Alfv\'enic turbulence,
we can better understand physical processes, e.g.~reconnection, particle acceleration, etc., in such media.

We describe the numerical methods in Section 2 and we present our results in Section 3.
We give discussions and summary in Section 4.

\section{Numerical Methods}
\subsection{Numerical Setups}
We solve the following system of equations in a periodic box of size $2\pi$:
\begin{equation}
   \frac{ \partial {\bf Q} }{ \partial t }
 + \frac{ \partial {\bf F} }{ \partial x^1 }
 =0,
\end{equation}
where
\begin{eqnarray}
{\bf Q}=(S_1,S_2,S_3,B_2,B_3), \\
 {\bf F}=(T_{11},T_{12},T_{13},-E_3,E_2), \\
 T_{ij}=-(E_iE_j + B_iB_j)+\frac{ \delta_{ij} }{2} (E^2+B^2), \\
 {\bf S}={\bf E}\times {\bf B}, \\
 {\bf E}=-\frac{1}{B^2} {\bf S}\times {\bf B}.
\end{eqnarray}
Here ${\bf E}$ is the electric field,
${\bf S}$ the Poynting flux vector, and we use units such that
the speed of light and $\pi$ do not appear in the equations
(see \citealt{Kom02} for details).

One can derive this system of equations from
\begin{eqnarray}
 \partial_{\mu} {}^{*}F^{\mu\nu}=0 \mbox{~~~(Maxwell's equation)}, \\
 \partial_{\mu} F^{\mu\nu}=-J^{\nu} \mbox{~~~(Maxwell's equation)}, \\
 \partial_{\mu} T^{\nu\mu}_{(f)}=0 \mbox{~~~(energy-momentum equation)}, \\
                F_{\nu\mu} u^{\mu}=0 \mbox{~~~(perfect conductivity)},
    \label{eq_cond}
\end{eqnarray}
where ${}^{*}F^{\mu\nu}$ is the dual tensor of the electromagnetic field,
$u^{\mu}$ the fluid four velocity, and $T^{\mu\nu}_{(f)}$ 
the stress-energy tensor of the electromagnetic field
\begin{equation}
 T^{\mu\nu}_{(f)}=F^{\mu}_{\alpha}F^{\alpha \nu} -\frac{1}{4}
                 (F_{\alpha \beta}F^{\alpha \beta})g^{\mu\nu},
\end{equation}
where $g^{\mu\nu}$ is the metric tensor and $F^{\alpha\beta}$ is
the electromagnetic field tensor.
We ignore the stress-energy tensor of matter.
We use flat geometry  and 
Greek indices run from 1 to 4.
One can obtain the force-free condition from 
Maxwell's equations and the energy-momentum equation
$\partial_{\mu} T^{\nu\mu}_{(f)}=-F_{\nu\mu}J^{\mu}=0$.
{}From Equation (\ref{eq_cond}), one can derive
\begin{eqnarray}
 {\bf E}\cdot {\bf B}=0, \\
 B^2-E^2 > 0.
\end{eqnarray}
In our simulations, the MHD condition ${\bf E} \cdot {\bf B}=0$ is enforced all the time.

We solve Equations (1)-(6) using a 
 Monotone Upstream-centered Schemes for Conservation Laws (MUSCL) type scheme with 
HLL fluxes
(Harten, Lax, van Leer 1983; in fact, in force-free MHD these fluxes reduce to
Lax-Friedrichs fluxes)
and monotonized central limiter (see Kurganov et al. 2001). 
The overall scheme is second-order accurate.
After updating the system of equations along the $x^1$ direction,
we repeat similar procedures for the $x^2$ and $x^3$ directions with
appropriate rotation of indexes.
Gammie, McKinney, \& T\'oth (2003) used a similar scheme for general relativistic MHD 
and Del Zanna, Bucciantini, \& Londrillo (2003) used
a similar scheme to construct a higher-order scheme for special
relativistic MHD.

While the magnetic field consists of the uniform background field and a
fluctuating field, ${\bf B}= {\bf B}_0 + {\bf b}$, the electric field
has only a fluctuating one.
The strength of
the uniform background field, $B_0$, is set to 1.
At $t=0$, no fluctuating fields are present.
We isotropically drive turbulence\footnote{
  We drive Alfv\'en waves only. 
  Nevertheless, our simulations naturally produce small amount of
  fast modes (see Cho 2005). 
  We ignore fast modes in this paper because their energy density is 
  small and they are passively cascaded by Alfv\'en modes \citep{ThoB98}.
}
 in the wave-number range $4\leq k \leq 6$.
We adjust the amplitude of forcing to maintain $b_+^2 \sim 1$ after saturation, where
the subscript `+' denotes dominant waves.
Therefore, we have
\begin{equation}
   \chi_+ \equiv \frac{ b_+ k_{\perp} }{ B_0k_{\|} } \sim 1
\end{equation}
after saturation.
Since the energy injection rates for the sub-dominant waves 
($\epsilon_-\equiv {\bf f}_- \cdot {\bf b}_-$) are equal to or less than those of
dominant waves ($\epsilon_+\equiv {\bf f}_+ \cdot {\bf b}_+$), where
${\bf f}$'s are forcing vectors,
we have $b_- \lesssim b_+$ and $\chi_- \lesssim 1$.
Simulation parameters are listed in Table 1.

\subsection{Test of the Code}
\begin{figure*}
\center
\includegraphics[angle=0,width=0.99\textwidth]{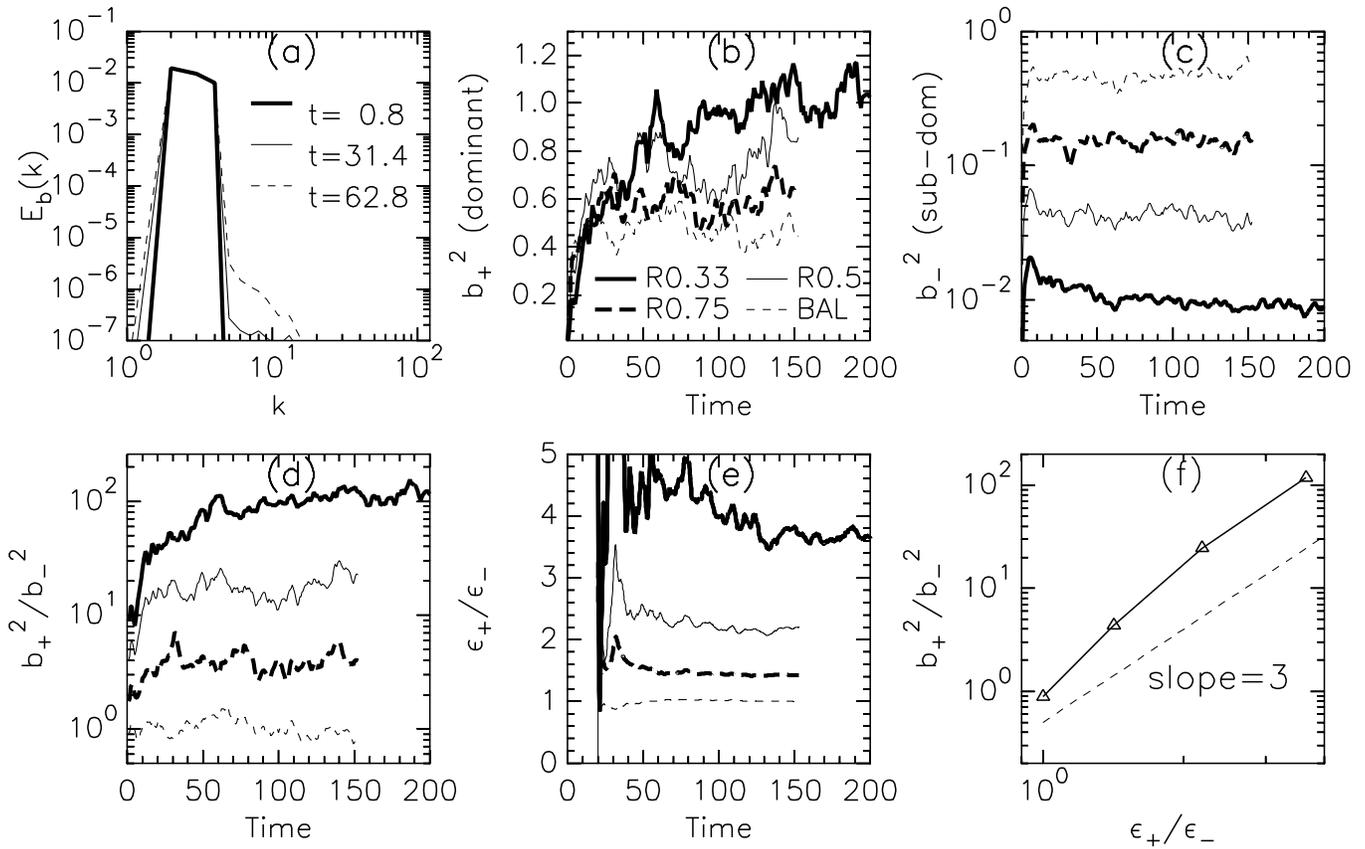}  
\caption{Energy densities and spectra. 
(a) Spectrum of waves moving in the same direction as a function of time.
    The spectrum does not change for a long time. 
(b) Time evolution of dominant modes. The lowest curve corresponds to the balanced turbulence (256-BAL).
     Note that $b_+^2 \sim B_0^2 =1$ for dominant waves.
(c) Time evolution of sub-dominant modes. From top to bottom, the degree of  imbalance increases.
(d) Time evolution of the value $b_+^2/b_-^2$.
(e) Time evolution of the value $\epsilon_+/\epsilon_-$; see the text for details.
(f) The relation between $<\epsilon_+/\epsilon_->$ and $<b_+^2/b_-^2>$.
Runs 256-BAL, 256-R0.75, 256-R0.5, and 256-R0.33 are used.
We use the same line convention for panels (b)-(e).
 }
\label{fig:1}
\end{figure*}

To check the stability of our code, we perform a simulation of relativistic Alfv\'en waves moving 
in the same direction.
Since Alfv\'en waves moving in one direction do not interact each other, their energy
spectrum should not change in time.
Indeed 
Figure \ref{fig:1}(a) confirms this: The initial energy spectrum (the thick solid line) does not show 
much change
even after t$\sim$63, which corresponds to $\sim$10 wave crossing times over the box size.

\begin{figure}
\center
\includegraphics[angle=0,width=0.95\columnwidth]{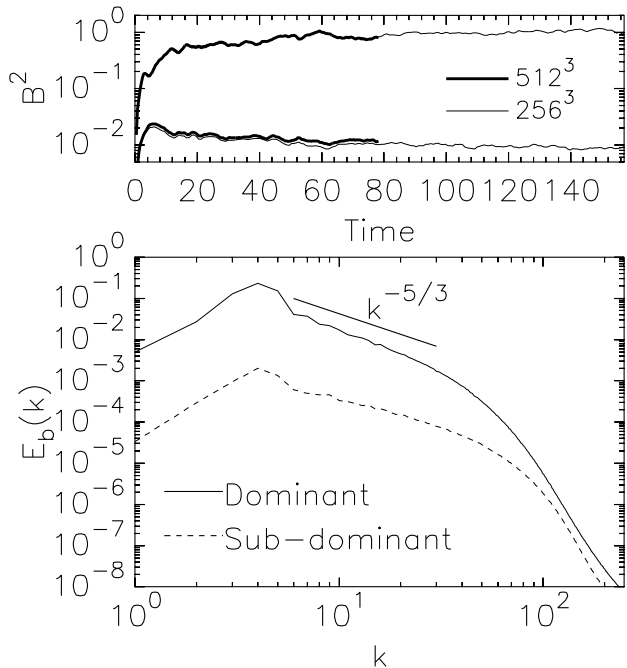}  
\caption{Results for 512-R0.33.
    Top panel: comparison between 256-R0.33 and 512-R0.33.
                              We plot time evolution of energy densities of dominant 
                              (upper curves) and sub-dominant (lower curves) waves.
   Bottom panel: energy spectra. The spectrum of the sub-dominant waves (dashed line) is shallower.
 }
\label{fig:2}
\end{figure}
\begin{figure*}[bh!]
\center
\includegraphics[width=0.95\textwidth]{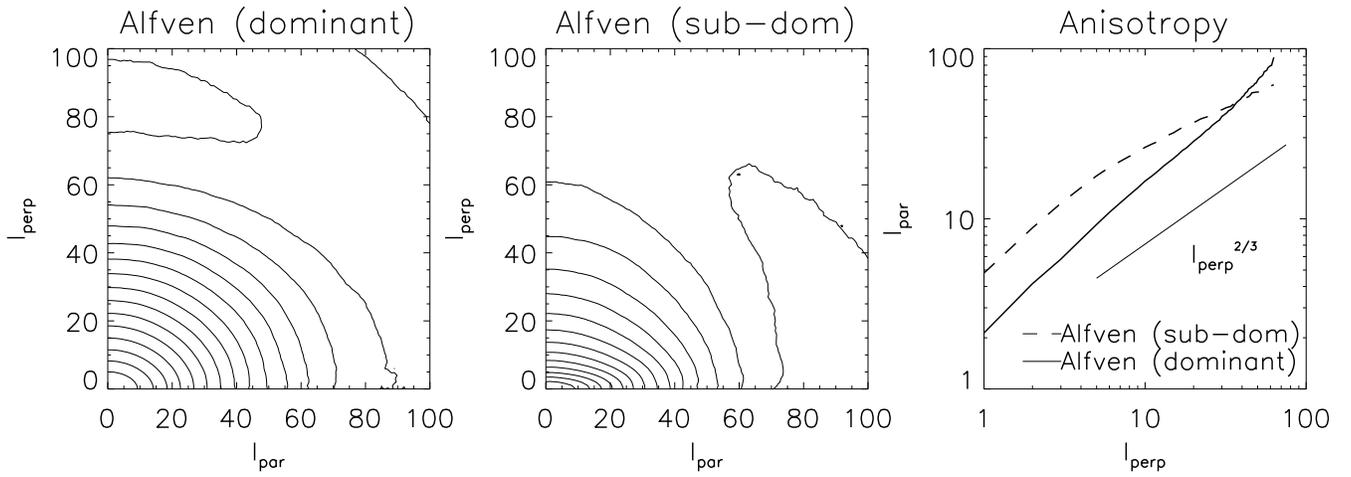}  
\caption{Anisotropy. Left panel: eddy shapes of dominant modes.
            Contours represent eddy shapes.
   Middle panel: eddy shapes of sub-dominant modes.
  Right panel: relation between semi-minor axes and semi-major axes of eddies
        (or, ``x intercepts'' and ``y intercepts'' of contours)
     {}(from 512-R0.33).
 }
\label{fig:3}
\end{figure*}

\begin{deluxetable}{lcl}
\tablecaption{Simulations \label{tab:table1}}
\tablewidth{0pt}
\tablehead{
\colhead{Run} & \colhead{Resolution} & \colhead{ $f_-/f_+$\tablenotemark{a} } 
}
\startdata
256-BAL     & 256$^3$ & 1 \\
256-R0.75 & 256$^3$ & 0.75\\
256-R0.5 & 256$^3$ & 0.5\\
256-R0.33 & 256$^3$ & 0.33\\
512-R0.33 & 512$^3$ & 0.33
\enddata
\tablenotetext{a}{Ratio of amplitudes of forcing. Subscripts `+' 
    and `-' denote dominant and sub-dominant modes, respectively.}
\end{deluxetable}


\section{Results}

\subsection{Energy Densities}
Figure \ref{fig:1}(b) shows time evolution of the energy densities of the dominant waves.
We have $b_+^2=b_-^2=0$ at $t=0$ and we drive the medium for $t>0$.
The energy densities of the dominant waves initially rise quickly  and reach saturation states.
The values of $b_+^2$ during saturation in those runs are between 0.5 and 1.0.
Since we drive turbulence isotropically, critical balance is roughly satisfied.
In general, the larger the imbalance, the slower the approach to the saturation state.
The largest imbalanced run (Run 256-R0.33) shows very slow approach to the saturation state.

Figure \ref{fig:1}(c) shows time evolution of energy densities of the sub-dominant waves.
{}From top to bottom, the degree of imbalance increases. The top curve corresponds to the balanced 
turbulence (Run 256-BAL) and the bottom curve to the largest imbalance (Run 256-R0.33).
Note that, in Run 256-R0.33, $b_-^2$ goes up very quickly for $0<t<5$ 
and then gradually goes down, which
may be due to the increase of $b_+$.

Figures \ref{fig:1}(d)  and (e) show time evolution of the ratio 
$b_+^2/b_-^2$ and $\epsilon_+/\epsilon_-$,
respectively. 
Figure \ref{fig:1}(d) clearly shows that
   the value of $b_+^2/b_-^2$ increases substantially as the degree of imbalance increases.
For $\epsilon_+/\epsilon_-$, we actually plot 
$\int_{t_0}^t \epsilon_+(t) dt / \int_{t_0}^t \epsilon_-(t) dt$, where
$t_0=20$.
Since different theories on imbalanced non-relativistic Alfv\'enic turbulence predict 
different relations between
$b_+^2/b_-^2$ and $\epsilon_+/\epsilon_-$, it will be useful to plot the relation for our simulations.
Figure \ref{fig:1}(f) shows the relation between the two ratios.
Roughly speaking, the ratio $b_+^2/b_-^2$ exhibits a power-law dependence on the ratio 
$<\epsilon_+/\epsilon_->$: $b_+^2/b_-^2 \propto (\epsilon_+/\epsilon_-)^n$ with $n>2$.

In Figure \ref{fig:1}, all simulations are performed on a grid of $256^3$ points.
The top panel of Figure \ref{fig:2}, which compares results of Runs 512-R0.33 and 512-R0.33, 
 implies that numerical resolution of $256^3$ would be enough for our current study.
Note that two runs have identical numerical set-ups except the numerical resolution ($256^3$ versus $512^3$).
The values of $b_+^2$ (upper curves) almost coincide, but the value of $b_-^2$ for $512^3$ is slightly
higher than that for $256^3$ (see lower curves).

\subsection{Spectra}
The bottom panel of Figure \ref{fig:2} shows energy spectra for Run 512-R0.33.
Although we have only about 1 decade of inertial range, we can clearly observe
that the spectral slopes for dominant and sub-dominant waves are different.
The spectrum of the dominant waves (upper curve) is slightly steeper than $k^{-5/3}$, while that of
the sub-dominant ones (lower curve) is a bit shallower than $k^{-5/3}$.

\subsection{Anisotropy}
In the presence of a strong mean magnetic field, structure of turbulence tends to elongate
along the direction of the mean field.
Therefore elongation of structures, or anisotropy, is an important aspect of  MHD turbulence. 
Both relativistic force-free and non-relativistic 
balanced Alfv\'enic turbulence are anisotropic.

Imbalanced non-relativistic Alfv\'enic turbulence is also anisotropic (e.g., \citealt{BerL08}).
Since interactions between eddies are very complicated in imbalanced Alfv\'enic turbulence,
it will be interesting to study anisotropy of imbalanced relativistic force-free MHD turbulence.

Figure \ref{fig:3} shows the shapes of eddies.
In the figure, we plot a contour diagram of the second-order
structure function for the magnetic field
in a local frame, which is aligned with the local mean magnetic field
${\bf B_L}$:
\begin{equation}
    \mbox{SF}_2(r_{\|},r_{\perp})=<|{\bf B}({\bf x}+{\bf r}) -
                 {\bf B}({\bf x})|^2>_{avg.~over~{\bf x}},
\end{equation}
where ${\bf r}=r_{\|} {\hat {\bf r}}_{\|} +r_{\perp} {\hat {\bf r}}_{\perp}$
and ${\hat {\bf r}}_{\|}$ and ${\hat {\bf r}}_{\perp}$ are unit vectors
parallel and perpendicular to the local mean field ${\bf B_L}$, respectively;
see Cho \& Vishniac (2000) and Cho et al.~(2002)
for the 
detailed discussion of the local frame.

The left and middle panels of Figure \ref{fig:3} show shapes of dominant and sub-dominant eddies,
respectively.
We can clearly see that the dominant eddies (left panel) are 
less anisotropic than the sub-dominant ones (middle
panel). If we plot the relation between perpendicular sizes of eddies 
(or, y intercepts of the contours; $\sim 1/k_\perp$)
 and
the parallel ones (or, x intercepts; $\sim 1/k_\|$), than we can see that
the dominant eddies show anisotropy weaker than $k_\| \propto k_\perp^{2/3}$ and 
the sub-dominant ones show anisotropy stronger than $k_\| \propto k_\perp^{2/3}$.

\subsection{Comparison with Non-Relativistic Theory and Simulations}
Our simulations are consistent with the theory and simulations of the imbalanced 
non-relativistic MHD turbulence (Beresnyak \& Lazarian 2008, 2009).
Indeed, the latter results are consistent with our finding 
of the relation between the ratio of the energy densities of the sub-dominant and dominant waves, 
their spectral slopes and their anisotropy. This is suggestive of a close relation
between the non-relativistic and relativistic turbulence
 and implies that the existing theories of
non-relativistic turbulence, e.g.~theories for magnetic reconnection, particle acceleration, etc., 
can be generalized for the relativistic limit.
This has not yet been done and, naturally, more theoretical/numerical research, especially with high numerical resolutions, for the relativistic case is necessary.

\section{Discussion and summary} 
Imbalanced turbulence is a generic incarnation of turbulence in the presence of sources and sinks of
turbulent energy. We know from the studies of non-relativistic imbalanced turbulence that its slower 
decay compared to the balanced one allows the energy transfer over larger distances and its transfer
to the balanced one due to parametric instabilities or the reflection of waves from density inhomogeneities
can result in local deposition of energy and momentum which provide many astrophysically
important consequences. The properties of imbalanced relativistic turbulence are important for many 
astrophysical settings
including the magnetosphere of pulsars, environments of gamma ray bursts and relativistic jets.

In this paper, we have studied imbalanced relativistic force-free MHD turbulence and
found the following results.
\begin{enumerate}
\item The magnetic spectrum of dominant waves is steeper than that of sub-dominant waves.
\item The dominant waves show  anisotropy weaker than 
      and  the sub-dominant waves show anisotropy  stronger than $k_\| \propto k_\perp^{2/3}$.
\item The energy density ratio $b_+^2/b_-^2$ is roughly proportional 
        to $(\epsilon_+/\epsilon_-)^n$, where $\epsilon$'s are energy injection rates and 
       $n>2$.
\end{enumerate}
All these results are consistent with the theory and simulations 
in Beresnyak \& Lazarian \citeyearpar{BerL08,BerL09}.
Therefore we can conclude that relativistic force-free MHD turbulence is indeed very similar to
its non-relativistic counterpart.

Our results imply that many results in non-relativistic Alfv\'enic turbulence can be
carried over to relativistic force-free MHD turbulence. For example, theories on 
magnetic reconnection (e.g.,~\citealt{LV99}), particle acceleration (e.g.,~\citealt{YL02}) 
and thermal diffusion (e.g.,~Cho et al. 2003) obtained 
in non-relativistic Alfv\'enic turbulence can 
also be applicable to relativistic force-free MHD turbulence.

The close similarity between the properties of non-relativistic and relativistic imbalanced turbulence
 found in this paper elucidates the nature of magnetic turbulence that preserves its properties in
both regimes irrespectively of whether turbulence is balanced or imbalanced. 
From the practical point of numerical studies, this allows us to 
test or double-check theories on non-relativistic Alfv\'enic turbulence 
using a completely different numerical scheme.

\acknowledgements
J.C.'s work is supported by the National R \& D Program through 
the National Research Foundation of Korea (NRF), 
funded by the Ministry of Education (No. 2011-0012081).
A.L. is supported by NSF grant AST 1212096, the Center for Magnetic Self-Organization and the
Vilas Associate Award. We thank the International Institute of Physics (Natal) for their
hospitality.

\end{document}